# Observation of Solar Radio Bursts Using E-CallistoSystem

J Adassuriya[1*], S Gunasekera[1], KPSC Jayaratne[2], C Monstein[3]

[1]Arthur C Clarke Institute for Modern Technologies, Katubedda, Moratuwa, Sri Lanka
[2]Department of Physics, University of Colombo, Colombo 03, Sri Lanka
[3]Institute for Astronomy, ETH Zurich, CH-8093 Zurich, Switzerland
adassuriya@accmt.ac.lk[*]

**ABSTRACT**
A CALLISTO system was set up at the Arthur C Clarke Institute and connected to the e-CALLISTO global network which observes the solar radio bursts in 24 hours. CALLISTO is the foremostobservation facility to investigate celestial objects in radio region in Sri Lanka. The system consists of the CALLISTO spectrometer and controlling software,logarithmic periodic antenna and pre-amplifier. CALLISTO spectrometer is able to detect solar radio bursts in the frequency range of 45MHz to870MHz with a channel resolution of 62.5kHz.The log-periodic antenna was designed for7dBi gain and achieved the voltage standing wave ratio, less than 1.5 which is acquired by the overall impedance ofthe antenna, 49.3Ω. The linear polarized antenna is pointing to zenith and the dipoles directed to north-south direction. The system detects solar radio emissions originated by solar flares and corona mass ejections. The radio bursts occurs as emission stripes in the radio spectra and classify from type I to V mainly on drift rate and band width. The system observed a type III solar radio burst on 5$^{th}$ July 2013 and a type II burst on 25$^{th}$ October 2013 which was originated by X1.7solar flare. The type II bursts characterize with narrow bandwidth and drift slowly from higher to lower frequencies while the main features of type III bursts are high drift rate and broad bandwidth.

## 1.0 INTRODUCTION

The solar flares and coronal mass ejections (CME's) are physical processes yet largely unknown but ongoing worldwide observation in different wavelength regions attemptingfor extensive understanding. In such attempts the exploration of solar events in metric wavelengths around 100 MHz and decimetric wavelength around 1 GHz are very vital because of the observationsthat can be done in all weather conditions and also for the availability of low cost instrumentations for measurements. One complication of radio observation is the interference of the artificial radio signals generated by numerous man made equipments. The natural radio sources in the solar system are the Earth, Jupiter and the most strongest, the Sun. The solar radio emission can be divided in to two parts, the disturbed and quiet sun emissions. The solar radio bursts are the emissions of disturbed sun when the sun is in its most active 11 – year solar cycle.

The solar radio bursts are classified into five different types based on frequency and temporal variations[1]. These bursts are energized by the different activities, solar flares, coronal mass ejections, solar noise storms which subsequently cause the geomagnetic storms, auroras and many atmospheric and magnetic disturbances. The impact of solar disturbances on satellite transmissions including Global Positioning System (GPS), power systems, radars and most of network systems is highly destructive due to its inability to predict them precisely. Therefore a world wide effort is launched by not only using earth based systems but also space based explorations to achieve the desired results. In such attempts, with the initiation of International Space Weather Initiative (ISWI), the CompactAstronomical Low-cost Low-frequency Instrument for Spectroscopy andTransportableObservatory (CALLISTO) has been introduced by Arnold O. Benz[2]and later the e-Callistonetwork was setup[3]which operates identical spectrographs different locations in the globe for 24 hour coverage.As a result of the International Heliophysical Year (IHY), an international program of scientific collaboration planned for 2007, more than 58 instruments in almost 32 locations were implemented.

**2.0 SYSTEM CONFIGURATION**

The eC48 Callisto spectrometer was deployed in October 2011at Arthur C Clarke Institute, Sri Lanka and started functioning from July 2012 and fully operating with the connection to e-Callisto system in May 2013. The station is located at $6^o$ 47' 37'' N, $79^o$53' 53'' E at an altitude 40m and the time zone +5.30 UT. The system consists of the Callisto spectrometer, log-periodic antenna with pre-amplifier and a Windows computer connected to the internet.

**2.1 Callisto Spectrometer**
The Callisto spectrometer eC48, based on a Philips CD1316LS/IV-3 tuner, a RS-232 shielded cable and a 12V power supply were donated by Institute of Astronomy of ETH Zurich, Switzerland. The Callisto was setup as a contribution to the International Space Weather Initiative (ISWI). A native Callisto spectrometer covers a frequency range from 45 MHz up to 870 MHz. Any other frequency range can be observed by switching in a heterodyne up- or down-converter.The programmable step size in frequency is 62.5 KHz while the radiometric band width Δfis about 300 KHz given by a ceramic band pass filter. The integration time Δt is 1 ms. Thus the rms noise σ of a Callisto is given by equation 1[4].

$$\sigma = \frac{T_{sys}}{\sqrt{\Delta f\ \Delta t}} = \frac{T_{sys}}{17.3} \quad (1)$$

Where $T_{sys}$ denotes the system temperature given by the preamplifier and the Callisto receiver. With standard components we can get down to about 80 K. Thus one σ will then be in the order of 5 K. For even better sensitivity expressed in temperature one would need a cooled preamplifier and/or longer integration and/or higher bandwidth. But for solar observations we need short integration time and small bandwidth to detect fine structures in the dynamic solar bursts.

Maximum sampling rate is 800 pixels per second, which the default is 200 channels per spectrum 4 sweeps per second. Any other combination is possible with the maximum of 400 channels can be stored in the internal memory (EEPROM) of Callisto. For high precision measurement in terms of timing Callisto can be clocked externally from a GPS system or an atomic clock signalwith 1 MHz TTL signal.

The intrinsic sensitivity of the logarithmic detector AD8307 is exactly 24.5mV/dBand it is maintaining the dynamic range of -110 dB to -50 dB. Dynamic range can be shifted by an electronic gain in the order of 40 dB to either cope with high level of interference on one hand or to go for highest sensitivity at locations with low rfi on the other hand. Callisto provides several different output formats. For debugging and quality assurance an ASCII formatted log-file is written all the time beginning every day at 00:00 UT. For interference analysis Callisto provides a function 'Save spectral overview'.This process steps through all possible channels from 45MHz up to 870MHz in steps of 62.5KHz leading to a total of 13200 channels. This file is also ASCII formatted providing two columns with frequency expressed in MHz and detector voltage expressed in mV. These files can be used to generate a spectral overview as shown in fig.1.Plot shown in fig. 1 is based on two measurements, first while observing the sky and a second one while observing a 50 Ω termination resistor at ambient temperature. Given the instrument bandwidth, physical temperature of the termination resistor the equivalent noise temperature and from that power flux can be determined (Rayleigh-Jeans-law in Jeans-region due to long wavelengths). For this measurement it is assumed that the antenna gain for radio frequency interference (rfi) is isotropic and thus 0 dB. Details can be found in ITU regulations[5]. For comparison the strongest natural radio sources (quiet Sun, Moon, Cygnus A and a 50 Ω termination) are also show in fig. 1. Dashed line represents the maximum limit for radio interference according to ITU-R RA.769 foreseen for VLBI observations.

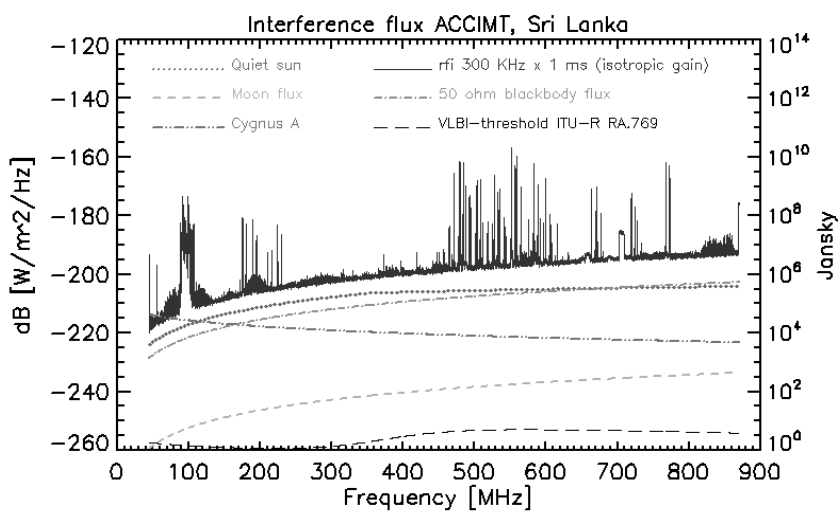

**Fig.1**: Spectral overview compared with natural sources

Data transfer is done either actively based on a tool called FTP-watchdog or data are collected from the server site with appropriate PERL scripts. Server and scripts are located at FHNW but maintained at ETH Zurich. The third file is the one which is used for solar science, it's a FIT file containing the observed intensity in a two dimensional array with x-axis expressed in UT, y-axis as frequency in MHz and intensity expressed in 'digits' in case of un-calibrated data.

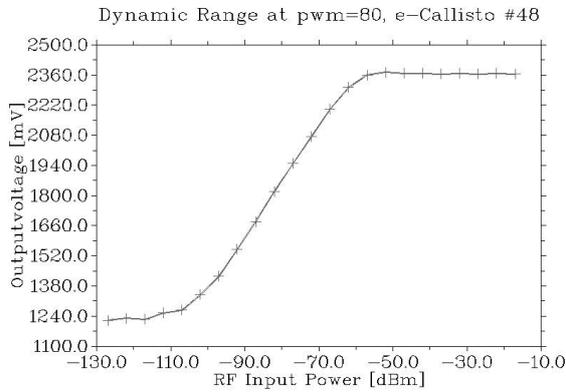
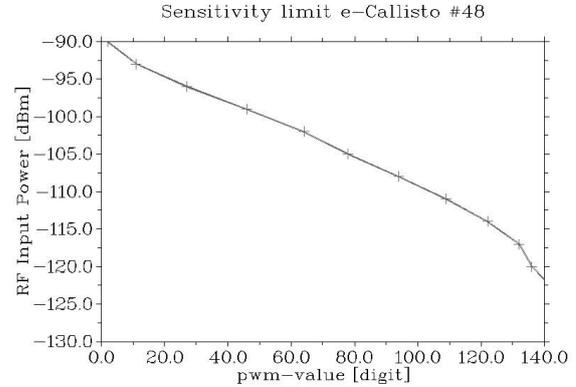

**Fig.2**: Dynamic range of Callisto measured during final check of the instrument

**Fig.3**: Sensitivity plot demonstrates that the minimum sensitivity can be shifted by 40 dB by applying a gain-control value (PWM).

Plots shown in figures 2 and 3 demonstrate quality measurements after production of the current instrument in the electronics laboratory of ETH Zurich. Left plot in fig. 2 proves the dynamic range of about 50 dB as well as the linearity of the detector within the range of operation. Saturation takes place when the antenna power exceeds -70 dBm. Right plot in fig. 3 proves the range for electronic gain control over a range of at least 30 dB. Electronic gain control is generated via a dc-voltage produced in the micro processor via a pulse width modulated (PWM) signal. In highly interfered observations site the system needs to be reduced to cope with high power signals which could led to cross modulation inside the receiver. On the other hand we can improve sensitivity at locations with low rfi to gain a few dB in sensitivity. This is all part of the configuration procedure described in the handbook[6].

**2.2 Logarithmic-Periodic antenna**
The log-periodic dipole antenna (LPDA) was designed and constructed by Arthur C Clarke Institute (ACCIMT). The basic arrangement of log-periodic antenna is two-wire line array elements of dipole antennas with 180°phase shift. Initially the design input parameters, theoretical gain of the LPDA, G=7dBi, nominal input resistance $R_o$=50Ω and the frequency range 45 MHz, the lowest ($f_l$) and 600 MHz, the highest ($f_n$) were set to achieve the voltage standing wave ratio (VSWR<1.5) and consequently the design constant ($\tau$) called the periodicity was chosen to be 0.822 andhence the relative spacing ($\sigma$), 0.149,was calculated usingoptimization equation2[7].

$$\sigma = 0.243\tau - 0.051 \qquad (2)$$

The bandwidth of the active region $B_{ar}$and structure bandwidth ($B_s$) are calculated from equations3 and 4[7].

$$B_{ar} = 1.1 + 7.7(1-\tau)^2 \frac{4\sigma}{1-\tau} \qquad (3)$$

$$B_s = B \times B_{ar} \qquad (4)$$

$$\text{where } B = \frac{f_n}{f_l}$$

The number of dipoles is mainly depended on the theoretical gain of the antenna. For 7 dBi gain and the frequency range of 45 - 600 MHz, the number of dipoles was determined by the equation 5[7], which results 18 dipoles.

$$N = 1 + \frac{\log B_s}{\log\left(\frac{1}{\tau}\right)} \qquad (5)$$

The longest dipole length was determined by the equation (6).

$$l_1 = \frac{1}{2}\frac{3 \times 10^8}{f_l} \qquad (6)$$

Subsequently the dipole lengths (*l*) and the relative spaces (*R*) in fig. 4 were calculated for 18 dipoles using the relation in equations 7 and 8.

$$l_{i+1} = \tau l_i \qquad (7)$$

$$R_{i+1} = \tau R_i \qquad (8)$$

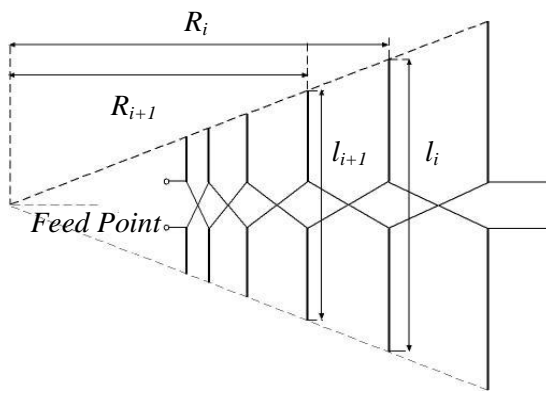

**Fig.4:** The arrangement of logarithmic periodic antenna.

In order to achieve the desired VSWR the length-to-diameter ratio (slimness factor) of the dipoles should be kept constant[7].Due to the standard diameter availability of the aluminum hollow tubes, it is impractical for the slimness factor to be kept constant. Therefore the diameter of the longer dipoles was chosen to be1.5cm, and 1.2cm for the medium sized dipoles, and 0.9cm for the shorter dipoles which results the average slimness factor S to be 74.

The designing of the antenna was entirely driven to achieve the lowest possible VSWR which can be obtained by matching the characteristic impedance of the antenna to the impedance of the Callisto receiver and the coaxial cable, RG-223/U. All the components of the system, antenna, pre-amplifier, Callisto receiver and RG-223/U cable should synchronize to 50Ω. The characteristic impedance of the overall antenna is give by equation 9[7].

$$Z_o = \frac{R_o^2 \sqrt{\tau}}{8 Z_{av} \sigma} + R_o \sqrt{\left(\frac{R_o \sqrt{\tau}}{8 Z_{av} \sigma}\right)^2 + 1} \qquad (9)$$

Where $Z_{av} = 120(\ln S - 2.25)$

For the average slimness factor S=74 and $R_o$=50Ω, the characteristic impedance led to the value $Z_o$=58.3Ω which is a good agreement with the desired value of 50Ω. 37x25mm aluminum channels were used for the booms and the two booms should separate by the distance given by equation 10[7].

$$d_f = b \cosh\left(\frac{Z_o}{120}\right) \qquad (10)$$

In the above equation, 'b' is the channel diameter which is approximated to b=1.18w, where w=25mm for the rectangular channel. Hence two booms were separated by 33mm using Perspex insulating sheets. The VSWR of the antenna for the entire frequency range, 45 – 870MHz, was able to be maintained less than 1.5 except only for the frequency range 45MHz to 127.5MHz.

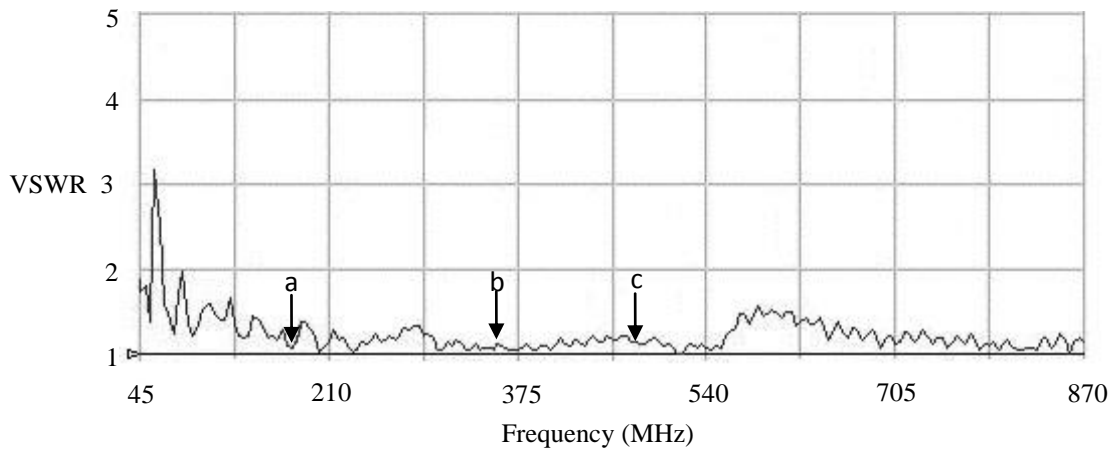

**Fig. 5:** Test results of VSWR of the LPDA for the entire frequency range 45 MHz to 870 MHz. Note that the marker points a, b and c are 176.8 MHz, 355.2 MHz, and 474.5 MHz and corresponding VSWR's are 1.0834, 1.0729 and 1.1564 respectively.

The impedance variation of the selected frequencies within the entire frequency range was obtained using the spectrum analyzer as shown in fig. 6. Due to the consideration of all aspects in designing phase, the average impedance of the LPDA, 49.3 Ω, was achieved which is almost equal to 50 Ω coaxial cable and the Callisto receiver. The impedance is highly deviated from the desired value for the frequencies 45 MHz, 57.5 MHz, and 61.5 MHz which reveals higher VSWR values in fig. 5. After some test observations the Callisto is set to operate in the ideal frequency range of 109 MHz to 452 MHz to minimize the strong radio interferences from FM bands and the most significant, the low VSWR (<1.5) ratio in the above frequency range.

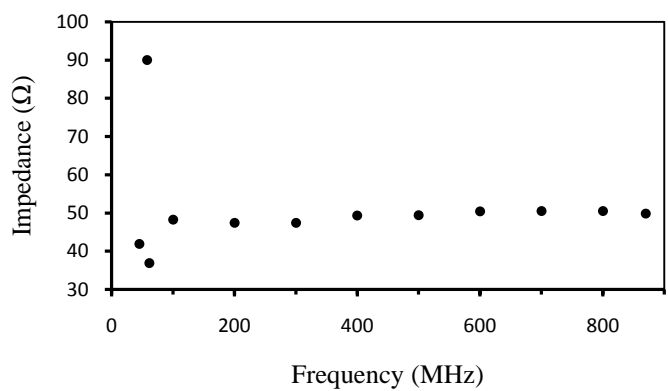

**Fig. 6:** The Impedance variation of the log-periodic antenna through the entire frequency range. Deviation of the impedance is higher at the low frequency range from 45MHz to 100MHz.

## 3.0 OBSERVATIONS

The continuous monitoring is being carried out from 01:00UT (6:30 local time) to 11:30 UT (17:00 local time) and data is uploaded to the FHNW server in the Flexible Image Transportation System (FITS) file every15 minutes. The system limitations, linear polarization, unavailability of a solar tracking system and the LPDA beam width of ±45° from the zenith,restrict observation only the strong radio bursts pointed to the earth. The gain parameter is set to a gain-control voltage via pulse width modulation (PWM) 150 and the number of frequency channels is set to the value 400 so that receiver sweeps the set frequency range two times per second.

### 3.1 First results
The solar radio bursts are classified as type I to type V based on the drift rate (df/dt), bandwidth and the structural parameters of the spectrum[1]. The system has detected, both type II and type III radio bursts at present as they are the most common radio bursts in the active sun. Fig. 7 shows the first result, a type III burst which occurred on 5 July 2013 due to a C2.2 class solar flare.

Type III bursts have relatively higher drift rate and broad band width compared to type II bursts[1] which is shown in fig. 8. A type II burst was observed by the system on 25 October 2013 originated by X1.7 solar flare. These observations were confirmed by detecting the similar radio bursts at Ooty system in India.

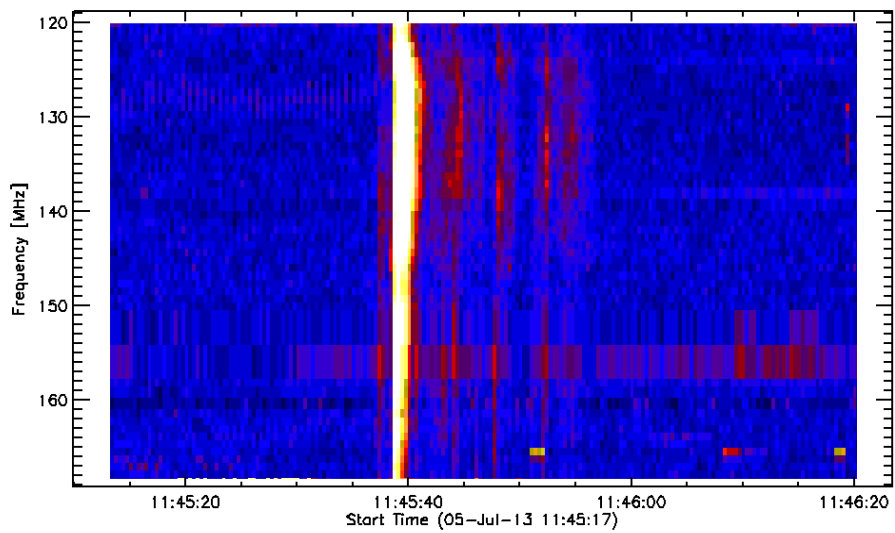

**Fig. 7**: Type III solar radio burst occurred on 2013 July 05 at 06:15:55 UT originated by C2.2 solar flare. At the time of observation the time stamp on the PC was wrongly set to local time instead of UT.

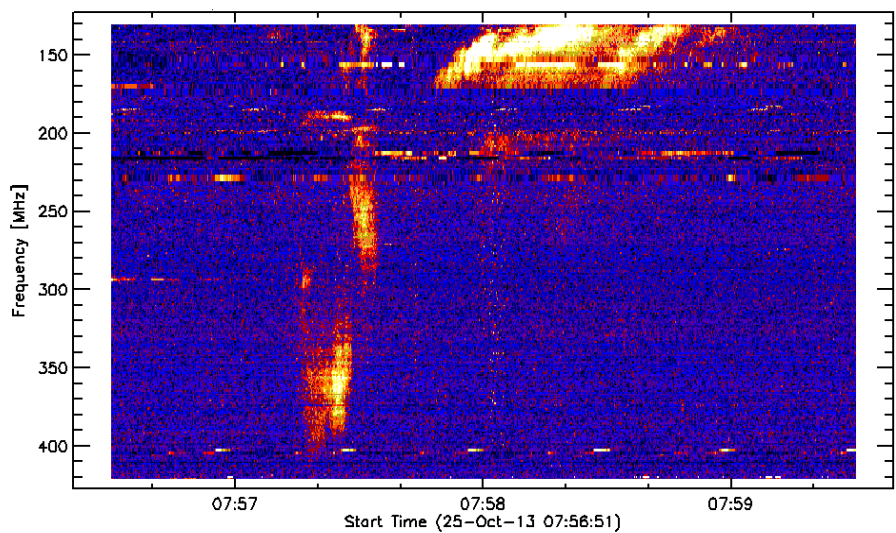

**Fig. 8**: Type II solar radio burst occurred on 2013 October 25 originated by X1.7 class solar flare at 07:53:00 UT. The radio part was delayed by about 4 minutes.

### 4.0 SUMMERY & DISCUSSION

In collaboration with ETH Zurich, the Arthur C Clarke Institute successfully constructed an e-CALLISTO station in Sri Lanka. The ETH Zurich provided the spectrometer plus software and ACCIMT constructed the LPDA antenna and pre-amplifier. The observation is being done covering 01:00 UT to 11:30 UT to observe the active sun in the 24$^{th}$ solar cycle. The system was tested for the best frequency range and set to operate avoiding the system anomalies and interferences of artificial radio bands. Interference level (rfi), taken from a single 15 minute FIT-file per location of the e-Callisto network confirms that the Sri Lankanstation is one of the best in low interference level.The bursts timings and classes were confirmed by latest solar flare event update by solar dynamic observatory (SDO) archive. The flare timings in fig. 7 and fig. 8 were obtained from SDO archive. The SDO observing the flares using the instrument, Atmospheric Image Assembly (AIA) at 131 $^{o}$A pass band region in extremeultraviolet and ultraviolet. Therefore the detection of radio part is not coincidedwith the timings of SDO observations. Nevertheless the observations are matched with the Callisto located at longitude 76$^{o}$ 42' 50'' at Ooty, India.